\documentstyle[12pt,epsf,epsfig,wrapfig]{article}
\begin{document}

\title{\hfill
\parbox[l]{4.5cm}{\normalsize ANL-HEP-CP-08-1} \\
{Spin-Orbit Dynamics from the Gluon Asymmetry}}
\author
{Gordon P. Ramsey\\Physics Dept., Loyola University Chicago and\\
HEP Division, Argonne National Lab \footnote{Talk given at the 
Spin Physics Symposium (DSPIN 2007), 3-9 September 2007, Dubna, 
Russia. Work supported by the USDOE Division of High Energy 
Physics, Contract DE-AC02-06CH11357. E-mail: gpr@hep.anl.gov}\\ \\
Yev Binder\\Physics Dept., Loyola University Chicago and \\ \\
Dennis Sivers\\Portland Physics Institute and\\Univ. of Michigan\\
}
\maketitle

\begin{abstract}
Determination of the orbital angular momentum of the proton is a difficult 
but important part of understanding fundamental structure. Insight can be 
gained from suitable models of the gluon asymmetry applied to the $J_z = 1/2$ 
sum rule. We have constrained the models of the asymmetry to gain possible 
scenarios for the angular momentum of the protonÕs constituents. Results and 
phenomenology for determining $L_z$ are presented. 
\end{abstract}

\section{Status of Proton Spin Structure}

For the past twenty years, much work has been done to understand the spin structure of the nucleons. 
There has been progress in determining the contribution of the lightest quarks to the spin, but there
is still uncertain knowledge about the gluon contribution. Transversity studies have contributed additional insight about quark dynamics, but little is known about the the orbital angular momentum
of the constituents.\cite{ji} This paper will summarize a project that provides a method of gaining insight into the nature of the orbital angular momentum of the nucleon constituents.

Recent experiments \cite{compass,hermes} have significantly lowered the measurement errors of the quark longitudinal spin contribution ($\Delta \Sigma$) to the proton. The COMPASS collaboration analysis quotes a result
\begin{equation}
\Delta \Sigma = 0.30\pm 0.01 (\mbox{stat}) \pm 0.02 (\mbox{evol}) ,\qquad \mbox{all data} \label{DSigC}
\end{equation}
while the HERMES collaboration analysis quotes a result
\begin{equation}
\Delta \Sigma = 0.330\pm 0.025 (\mbox{exp}) \pm 0.011 (\mbox{th})\pm 0.028 (\mbox{evol}),\qquad \mbox{all data} \label{DSigH}.
\end{equation}
These groups and others \cite{rhic} have been working on providing a significant measure of the proton's spin weighted gluon density,
\begin{equation}
\Delta G(x,t)\equiv G_{++}(x,t)-G_{+-}(x,t),
\end{equation}
where $x$ is the Bjorken scaling variable and $t\equiv \log(\alpha_s(Q_0^2)/\log(\alpha_s(Q^2))$ is the $Q^2$ evolution variable. The combination of these measurements is summarized in terms of the 
$J_z={1\over 2}$ sum rule:
\begin{equation}
J_z={1\over 2}\equiv {1\over 2}\Delta \Sigma+\Delta G+L_z. \label{JSR}
\end{equation}
Here $\Delta \Sigma=\int^1_0\>dx\Delta q(x,t)$ and $\Delta G=\int^1_0\>dx\Delta G(x,t)$ are the 
projections of the spin carried by all quarks and the gluons on the $z$-axis, respectively. Also $L_z$ is the net $z$-component of the orbital angular momentum of the constituents. We do not attempt to separate the flavor components of $L_z$ within the sum rule. \\

\section{Modeling the Gluon Asymmetry}

Experimental groups at the COMPASS, HERMES and RHIC collaborations are measuring both the gluon polarization and the asymmetry, $A\equiv \Delta G/G$ to determine the gluon polarization 
\cite{compass, hermes, rhic}. Since there is no suitable theoretical model for $\Delta G$, we have 
devised a way to model the asymmetry, $A(x,t)$ to gain insight into the structure of $\Delta G$. This, 
coupled with the $J_z={1\over 2}$ sum rule can then shed light on the nature of the orbital angular
momentum of the constituents, $L_z$. To model $A(x,t)$, we write the polarized gluon asymmetry using the decomposition
\begin{equation}
A(x,t)\equiv \Delta G/G=A_0(x)+\epsilon(x,t), \label{Adef}
\end{equation}
where
\begin{equation}
A_0(x)\equiv \Bigl[({{\partial \Delta G}\over {\partial t}})/
({{\partial G}\over {\partial t}})\Bigr] \label{A0def}
\end{equation}
is a scale invariant calculable reference form \cite{gpr}. Here 
$\epsilon(x,t)$ represents the difference between the calculated and gauge-invariant asymmetry. Since $\Delta G$ is unknown, a 
useful form is to write equation (\ref{Adef}) as
\begin{equation}
\Delta G = A_0(x)\>G(x,t)+\Delta G_{\epsilon}(x). \label{DG}
\end{equation}
Although the quantity $\Delta G_{\epsilon}(x)$ is not a physical parameter, it allows the theoretical
development of the calculable quantity, $A_0$. Once an asymmetry is generated from equations 
(\ref{A0def}) and (\ref{DG}), the gauge-invariant quantity 
$A(x,t)$ can be compared to data. Thus, each Ansatz for 
$\Delta G_{\epsilon}(x)$ gives a corresponding form for 
$\Delta G$ and a parametrization for $L_z$. These can be compared 
to existing data to provide a range of suitable models for these 
contributions. 

With the definition for the asymmetry in equation (\ref{A0def}), the DGLAP equations can then be used to 
evaluate the evolution terms on the right side. 
\begin{equation}
A_0=\Biggl[{{\Delta P_{Gq} \otimes \Delta q+\Delta P_{GG}\otimes 
\Delta G}\over {P_{Gq} \otimes q+P_{GG}\otimes G}}\Biggr]. \label{A00}
\end{equation}
The polarized gluon distribution in the numerator of equation (\ref{A00}) is replaced by 
$\Delta G\equiv A_0\cdot G+\Delta G_{\epsilon}$. For certain unpolarized distributions, there are points 
at which the denominator vanishes. To avoid this, we write equation (\ref{A00}) as: 
\begin{eqnarray}
{{\partial{\Delta G}}\over {\partial{t}}}&=&(2/\beta_0) \Bigl[\Delta P_{gq}^{LO}\otimes \Delta q+\Delta P_{gg}^{LO}\otimes (A_0\cdot G+\Delta G_{\epsilon})\Bigr]  \\ \nonumber
 &=&A_0\cdot{{\partial{G}}\over {\partial{t}}} \\
 &=&(2/\beta_0) A_0 \Bigl[P_{gq}^{LO}\otimes q+P_{gg}^{LO}\otimes G\Bigr]. \nonumber \label{ADGe}
\end{eqnarray}
The NLO form is essentially the same as equation (9) with the splitting functions $P^{LO}$
replaced with their NLO counterparts. The quark and gluon unpolarized distributions are CTEQ5 and
the polarized quark distributions are a modified GGR set. \cite{GGR} 

There are constraints on $A_0(x)$ that must be imposed to satisfy the physical behavior of the gluon
asymmetry, $A(x)$. These are:
\begin{itemize}
\item positivity: $|A_0(x)| \le 1$ for all x, and
\item endpoint values: $A_0(0)=0$ and $A_0(1)=1$
\end{itemize}
Note that the constraint of $A_0\to 1$ is built in to satisfy the assumption that the large $x$ 
parton distributions are dominated by the valence up quarks in the proton. The 
convolutions are dominated by the quark terms, which force the asymmetry to unity as $x\to 1$. 
To investigate the possible asymmetry models, we use a parameterization for $A_0$ in the form
\begin{equation}
A_0\equiv Ax^{\alpha}-(B-1)x^{\beta}+(B-A)x^{\beta+1}, \label{A0form}
\end{equation}
which automatically satisfies the constraints that $A_0(0)=0$ and $A_0(1)=1$. Once a parametrization 
for $\Delta G_{\epsilon}(x)$ is chosen, equation (9) is used to determine the parameters in 
equation (\ref{A0form}). 

\section{Results and Conclusions}

The models for $\Delta G_{\epsilon}(x)$ that led to asymmetries that satisfied these constraints were all
in the range $|\int^1_0 \Delta G_{\epsilon} dx|\le 0.25$, with positive and negative values included. 
Larger values of $\Delta G_{\epsilon}$ violate one or both of the constraints. A representative sample of 
models that satisfy the constraints are listed in Table 1.
\begin{table}[htdp]
\caption{Gluon Asymmetry Parameters}
\begin{center}
\begin{tabular}{||c|c|c|c||}
\hline
$\Delta G_{\epsilon}$ & $\int _0^1 \Delta G_{\epsilon} dx$ & $A_0$ & $\int _0^1 \Delta G dx$ \\
\hline
\hline
$0$ & $0$ & $3x^{1.5}-3x^{2.2}+x^{3.2}$ & 0.18 \\
\hline
$2(1-x)^7$ & $0.25$ & $4x^{1.6}-4x^{2.1}+x^{3.1}$ & 0.42 \\
\hline
$-2(1-x)^7$ & $-0.25$ & $1.75x^{1.1}-1.5x^{2.1}+0.75x^{3.1}$ & 0.01 \\
\hline
$-90x^2(1-x)^7$ & $-0.25$ & $3.5x^{1.3}-4.5x^{2.2}+2x^{3.2}$ & 0.05 \\
\hline
$9x(1-x)^7$ & $0.125$ & $3.75x^{1.4}-3x^{1.6}+0.25x^{2.6}$ & 0.29 \\
\hline
$-9x(1-x)^7$ & $-0.125$ & $3.25x^{1.4}-3.75x^{2.2}+1.5x^{3.2}$ & 0.11 \\
\hline
$4.5x(1-x)^7$ & $0.0625$ & $2x^{0.9}-1.5x^{1.2}+0.5x^{2.2}$ & 0.37 \\
\hline
$-4.5x(1-x)^7$ & $-0.0625$ & $2.25x^{1.1}-2.25x^{1.9}+x^{2.9}$ & 0.23 \\
\hline
\end{tabular}
\end{center}
\label{default}
\end{table}

Note that the integrals for $\Delta G$ are all positive, ranging from about 0.01 to 0.42. The models that 
gave negative values for these integrals did not agree with the existing asymmetry data, reported at this 
workshop to be:
\begin{itemize}
\item $\Delta G/G=0.016\pm 0.058\pm 0.055$ at $x=0.09$ from COMPASS, $Q^2>1$ GeV$^2$
\item $\Delta G/G=0.060\pm 0.31\pm 0.06$ at $x=0.13$ from COMPASS, $Q^2<1$ GeV$^2$
\item $\Delta G/G=0.078\pm 0.034\pm 0.011$ at $x=0.204$ from HERMES, factorization method
\item $\Delta G/G=0.071\pm 0.034\pm 0.010$ at $x=0.222$ from HERMES, approximate method.
\end{itemize}
The models in Table 1 that are within one $\sigma$ of the preliminary data stated above are in the third, 
fourth and sixth rows, respectively. Plots of the full asymmetry are shown in Figure 1. None of the models in Table 1 are ruled 
out by the data since they fall within two $\sigma$ of the data for our values of $Q^2>1$ GeV$^2$. All of these models except for 
the fourth row in Table 1 (impulses in figure 1) generate total 
asymmetries $A(x,t=0)$ that are close to $A(x)=x$. Ironically, 
early assumptions of the polarized gluon assumed this 
functional form as a naive estimate to the asymmetry. Next-to-leading order corrections to these 
asymmetries tend to bring them less positive, but with the same general shape. A full set of viable asymmetries will be presented
in an upcoming paper. \cite{brs}

\begin{figure}[htbp]
   \begin{center}
   \rotatebox{270}{\resizebox{8cm}{!}{\includegraphics{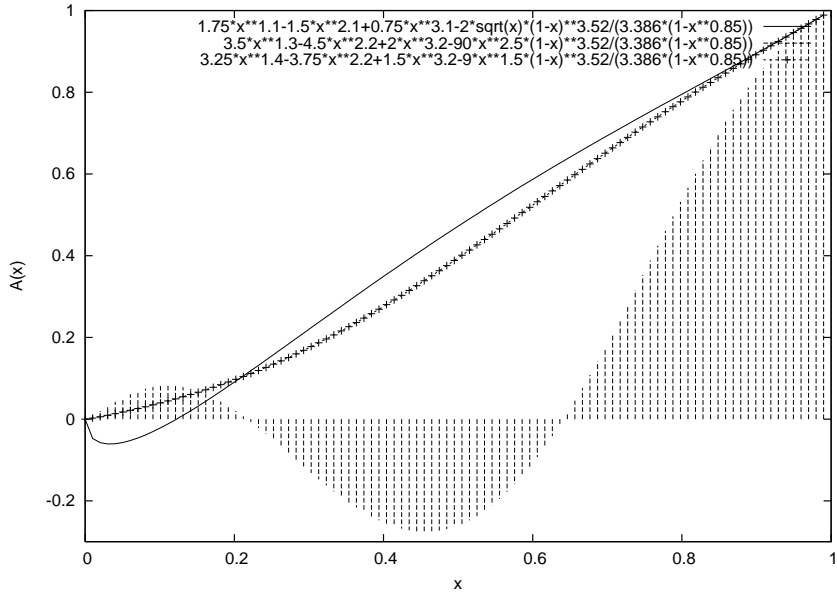}}} 
   \caption{The gluon asymmetries most closely in agreement with data. Solid line, impulses and 
   linespoints represent the models in rows 3, 4 and 6 of Table 1 respectively.}
   \label{ramsey_fig1}
   \end{center}
\end{figure}

Using the data on $\Delta \Sigma$ in Section 1, the relation between $<\Delta G>$ and $<L_z>$ can be written as: 
\begin{equation}
<\Delta G>=0.35\>-<L_z>\pm 0.02. \label{DGLz}
\end{equation}
The three models of the asymmetry that agree most closely with existing data give values of $\Delta G$ in the approximate range of $0\to 0.11$. Thus, the existing data with equation \ref{DGLz} imply the 
approximate relation $0.24\le L_z\le 0.35\pm 0.02$. Thus, the contribution of the orbital motion of the
constituents to the proton spin may be comparable to the total quark contribution. A recent lattice calculation of the 
contribution of the quark orbital motion to the proton spin 
($L_z^q$) is consistent with zero. \cite{lattice} Thus, the 
gluonic orbital motion appears to provide the majority contribution to $L_z$ in the $J_z={1\over 2}$ sum rule.
It is clear that future measurements of $\Delta G$ and 
$\Delta G/G$ must be made in a wider kinematic range of $x$ and 
$Q^2$ with improved precision to better specify the appropriate 
model of the asymmetry and to extract the $x$ and $Q^2$
dependence of the orbital angular momentum of the constituents.

\end{document}